\documentclass[aps,prb,twocolumn,showpacs,preprintnumbers,amsmath,amssymb,superscriptaddress]{revtex4}%

\usepackage{graphicx}%
\usepackage{dcolumn}
\usepackage{amsmath}
\usepackage{color}

\makeatletter
\def\btt#1{\texttt{\@backslashchar#1}}%
\DeclareRobustCommand\bblash{\btt{\@backslashchar}}%
\makeatother

\begin{document}

\preprint{PREPRINT (\today)}

\title{ Muon-spin rotation measurements of the penetration
depth of the Mo$_3$Sb$_7$ superconductor}
\author{R.~Khasanov}
 \email{rustem.khasanov@psi.ch}
 \affiliation{Laboratory for Muon Spin Spectroscopy, Paul Scherrer
Institut, CH-5232 Villigen PSI, Switzerland}
\author{P.W.~Klamut}
\affiliation{Institute of Low Temperature and Structure Research
of the Polish Academy of Sciences, Ok\'{o}lna 2, 50-422
Wroc{\l}aw, Poland }
\author{A.~Shengelaya}
 \affiliation{Physics Institute of Tbilisi State University,
Chavchavadze 3, GE-0128 Tbilisi, Georgia}
\author{Z. Bukowski}
 \affiliation{Institute of Low Temperature and Structure Research
of the Polish Academy of Sciences, Ok\'{o}lna 2, 50-422
Wroc{\l}aw, Poland }
 \affiliation{Laboratory for Solid State Physics, ETH Z\"urich, CH-8093 Z\"urich,
Switzerland}
\author{I.M.~Savi\'c}
 \affiliation{Faculty of Physics, University of Belgrade, 11001
Belgrade, Serbia and Montenegro}
\author{C.~Baines}
\affiliation{Laboratory for Muon Spin Spectroscopy, Paul Scherrer
Institut, CH-5232 Villigen PSI, Switzerland}
\author{H.~Keller}
\affiliation{Physik-Institut der Universit\"{a}t Z\"{u}rich,
Winterthurerstrasse 190, CH-8057 Z\"urich, Switzerland}

\begin{abstract}
Measurements of the magnetic field penetration depth $\lambda$ in
superconductor Mo$_3$Sb$_7$ ($T_c\simeq2.1$~K) were carried out by
means of muon-spin-rotation. The absolute values of $\lambda$, the
Ginzburg-Landau parameter $\kappa$, the first $H_{c1}$ and the
second $H_{c2}$ critical fields at $T=0$ are
$\lambda(0)=720(100)$~nm, $\kappa(0)=55(9)$,
$\mu_0H_{c1}(0)=1.8(3)$~mT, and $\mu_0H_{c2}(0)=1.9(2)$~T. The
zero temperature value of the superconducting energy gap
$\Delta(0)$ was found to be 0.35(1)~meV corresponding to the ratio
$2\Delta(0)/k_BT_c=3.83(10)$. At low temperatures
$\lambda^{-2}(T)$ saturates and becomes constant below $T\simeq
0.3T_c$, in agreement with what is expected for $s-$wave BCS
superconductors. Our results suggest that Mo$_3$Sb$_7$ is a BCS
superconductor with the isotropic energy gap.
\end{abstract}

\maketitle


Recently, the attention was devoted  to Mo$_3$Sb$_7$. This
compound was originally discovered more than forty years ago
\cite{Jensen66,Hulliger66} and only recently was found to become a
type-II superconductor with the transition temperature
$T_c\simeq2.1$~K.\cite{Bukowski02} The properties of Mo$_3$Sb$_7$
in a superconducting state are rather unusual.
Specific heat, resistivity and magnetic susceptibility experiments
of Candolfi {\it et al.}\cite{Candolfi07} suggest that
Mo$_3$Sb$_7$ can be classified as a coexistent superconductor --
spin-fluctuating system.
As discussed in Ref.~\onlinecite{Candolfi07},  factoring in the
effect of spin fluctuations leads to renormalized values of the
electron-phonon coupling constant and the Coulomb pseudopotential,
which, being substituted to the McMillan expression, lead to $T_c$
between 1.4~K and 2.0~K. This is substantially closer to the
experimentally observed  $T_c\simeq 2.1$~K than $T_c\approx10$~K,
which would be obtained without taking into account the effect of
spin fluctuations.\cite{Candolfi07}

There is currently no agreement on the symmetry of the order
parameter of Mo$_3$Sb$_7$. The recent specific heat experiments of
Candolfi {\it et al.}\cite{Candolfi08} suggest that the order
parameter in Mo$_3$Sb$_7$ is of conventional $s-$wave symmetry.
Tran {\it et al.}\cite{Tran08}, based again on the results of
specific heat measurements, have reported the presence of two
isotropic $s-$wave like gaps with $2\Delta_1(0)/k_BT_c=4.0$ and
$2\Delta_2(0)/k_BT_c=2.5$ [$\Delta(0)$ is the zero-temperature
value of the superconducting energy gap]. In contrast, Andreev
reflection measurements of Dmitriev {\it et
al.}\cite{Dmitriev06,Dmitriev07a,Dmitriev07b} reveal  that the
superconducting gap is highly anisotropic. The maximum to the
minimum gap ratio was estimated to be $\Delta_{max}/\Delta_{min}\simeq
40$ and $s+g-$wave symmetry of the order parameter was proposed in
a qualitative analysis.\cite{Dmitriev07a,Dmitriev07b}

The symmetry of the superconducting order parameter can be probed
by measurements of the magnetic penetration depth $\lambda$. A
fully gaped, isotropic pairing state produces a thermally
activated behavior leading to an almost constant value of the
superfluid density $\rho_s\propto\lambda^{-2}$ for
$T\lesssim0.3T_c$.\cite{Khasanov05,Prozorov06}
Presence of line nodes in the gap leads to a continuum of  low
laying excitations, which result in a linear $\lambda^{-2}(T)$ at
low temperatures.\cite{Hardy93,Kadono04}
In two-gap superconductors with highly different gap to $T_c$
ratios the inflection point in $\lambda^{-2}(T)$ is generally
present.\cite{Carrington03,Khasanov07_La214,Khasanov07_Y123}

In this paper, we report the study of the magnetic field
penetration depth in superconductor Mo$_3$Sb$_7$ by means of
muon-spin rotation. Measurements were performed down to 20~mK in a
series of fields ranging from 0.02~T to 0.2~T. Our results are
well explained assuming conventional superconductivity with the
{\it isotropic} energy gap in agreement with the recent
specific heat experiments of Candolfi {\it et
al.}\cite{Candolfi08} The zero-temperature value of the gap was
found to be $\Delta(0)=0.35(1)$~meV corresponding to the ratio
$2\Delta(0)/k_BT_c=3.83(10)$.


The Mo$_3$Sb$_7$ single-crystal samples were grown through
peritectic reaction between Mo metal and liquid
Sb.\cite{Bukowski02}
The transverse field muon-spin rotation (TF-$\mu$SR) experiments
were performed at the $\pi$M3 beam line at Paul Scherrer Institute
(Villigen, Switzerland). For our experiments the ensemble of some
sub- and millimeter size single crystals were mounted onto the
silver plate to cover the area of approximately 50~mm$^2$. The
silver sample holder was used because it gives a nonrelaxing muon
signal and, hence, only contributes as temperature independent
constant background. The crystals were oriented so that the
magnetic field was preferably applied along the 001
crystallographic direction. The Mo$_3$Sb$_7$ samples were field
cooled from above $T_c$ down to $\simeq20$~mK in magnetic fields
ranging from 20~mT to 0.2~T. A full temperature scan (from 20~mK
up to 2.5~K) was performed in a field of $\mu_0H=0.02$~T.

\begin{figure}[htb]
\includegraphics[width=1.0\linewidth]{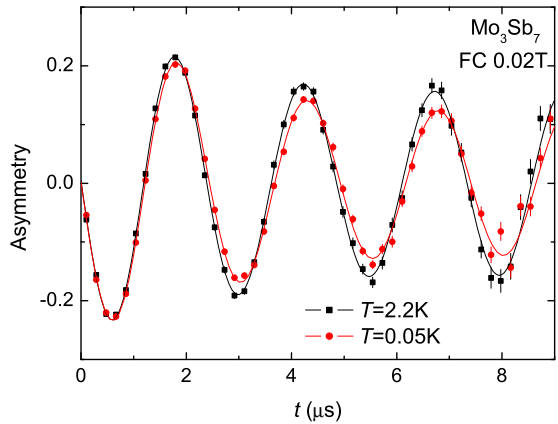}
 \vspace{-1.0cm}
\caption{(Color online) Transverse-field muon-spin precession
signals from Mo$_3$Sb$_7$ obtained in $\mu_0H=0.02$~T above the
transition temperature ($T=2.2$~K -- black squares) and after
field-cooling the sample below $T_c$ ($T=0.05$~K -- red circles).
The solid lines correspond to the fit by means of
Eq.~\ref{eq:Analysis}. For visualization purpose the apparent
precession frequencies are modified from the actual precession
frequencies by the use of a rotating reference frame.}
 \label{fig:Time-Spectra}
\end{figure}

Figure~\ref{fig:Time-Spectra} shows the muon-spin precession
signals in $\mu_0H=0.02$~T above ($T=2.2$~K) and below
($T=0.05$~K) the superconducting transition temperature  of
Mo$_3$Sb$_7$. The difference in the relaxation rate above and
below $T_c$ is due to the well-known fact that type-II
superconductors exhibit a flux-line lattice leading to spatial
inhomogeneity of the magnetic induction. As is shown by
Brandt,\cite{Brandt88,Brandt03} the second moment of this
inhomogeneous field distribution is related to the magnetic field
penetration depth $\lambda$ in terms of $\langle \Delta
B^{2}\rangle\propto\sigma_{sc}^2\propto\lambda^{-4}$.

The analysis of TF-$\mu$SR data was carried out in the time-domain
by using the following functional form:\cite{Adroja05}
\begin{eqnarray}
A(t)&=&A_0 \exp\left[-\frac{\sigma_{nm}^2+\sigma_{sc}^2}{2}\
t^2\right]
\cos(\gamma_\mu  B_{int}\ t+\phi) \nonumber\\
&&+A_{bg}\cos(\gamma_\mu  B_{ext}\ t+\phi),
 \label{eq:Analysis}
\end{eqnarray}
where the first term denotes the contribution from the
Mo$_3$Sb$_7$ sample  and the second term is the background
contribution from the Ag sample holder. $A_0$ and $A_{bg}$ are
the initial asymmetries arising from the sample and the
background, $B_{int}$ and $B_{ext}$ are the internal filed inside
the sample and the applied external field seen in the Ag backing plate, $\gamma_{\mu} =
2\pi\times135.5342$~MHz/T is the muon gyromagnetic ratio, and
$\phi$ is the initial phase. $\sigma_{sc}$ and $\sigma_{nm}$ are
the muon-spin relaxation rates caused by the nuclear moments and
the additional component appearing below $T_c$ due to nonuniform
field distribution in the superconductor in the mixed state.

The analysis was carried out as follows. First, the muon-time
spectra  were fitted by means of Eq.~(\ref{eq:Analysis}) with all
the parameters free. Then, the ratio $A_{0}/A_{bg}=0.975$ was
obtained as the mean value in the temperature interval of
$0.02$~K$\leq T \leq 1.5$~K. After that, by keeping the ratio
$A_{bg}/A_0$ fixed, the nuclear moment contribution $\sigma_{nm}$
was estimated by analyzing the experimental data above 2.1~K,
({\it i.e.} at $T>T_c$, where $\sigma_{sc}=0$). The final analysis
was made with $A_{0}/A_{bg}$, $\sigma_{nm}$ and $B_{ext}$ as fixed, and
$B_{int}$  and $\sigma_{sc}$ as free parameters. This
allows to reduce the number of the fitting parameters and, as a
consequence, to increase the accuracy in determination of
$\sigma_{sc}$.


Within BCS theory the zero-temperature magnetic penetration depth [$\lambda(0)$]
of isotropic superconductor does not depend on the magnetic field.
Contrary, the nonlocal and the nonlinear response of the
superconductor  with nodes in the gap to the magnetic field, as
well as, the faster suppression of the contribution of the smaller
gap to the total superfluid density in a case of two-gap
superconductor leads to the fact that $\lambda(0)$, evaluated from
$\mu$SR experiments, is magnetic field dependent and it increases
with increasing field.\cite{Amin00,Landau07} Such behavior was
observed in various hole-doped
cuprates,\cite{Kadono04,Sonier00,Khasanov07_field-effect} and the
double-gap NbSe$_2$\cite{Sonier00,Fletcher07} and MgB$_2$
superconductors.\cite{Serventi04,Angst04}

\begin{figure}[htb]
\includegraphics[width=1.0\linewidth]{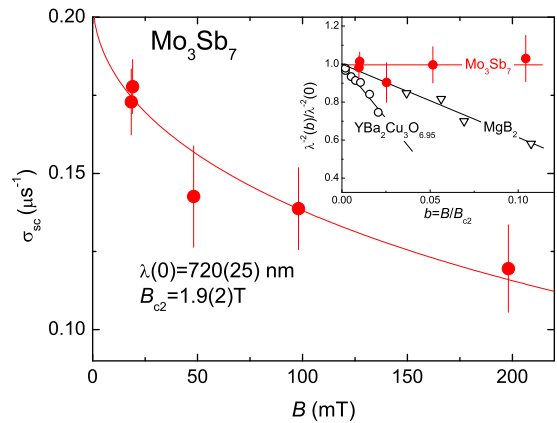}
 \vspace{-1.0cm}
\caption{(Color online) Field dependence of the $\mu$SR
depolarization  rate $\sigma_{sc}$ measured at $T=20$~mK in
Mo$_3$Sb$_7$ sample. The solid line corresponds to a fit of
Eq.~(\ref{eq:sigma_vs_h}) to the experimental data with the
parameters shown in the figure. The inset shows the dependence of
$\lambda^{-2}$ normalized to its value at $B=0$ on the reduced
magnetic field $b=B/B_{c2}$ for Mo$_3$Sb$_7$, the two-gap
superconductor MgB$_2$ from Ref.~\onlinecite{Angst04}, and the
hole-doped cuprate superconductor YBa$_2$Cu$_3$O$_{6.95}$ from
Ref.~\onlinecite{Sonier00}.}
 \label{fig:lambda_vs_H_muSR}
\end{figure}

Consideration of the ideal vortex lattice (VL) of an
isotropic  $s-$wave superconductor within Ginsburg-Landau approach
leads to the following expression for the magnetic field
dependence of the second moment of the magnetic field
distribution:\cite{Brandt03}
\begin{eqnarray}
 \sigma_{sc}[\mu {\rm s}^{-1}]=a\times
(1 - B/B_{c2}) \nonumber \\
 \left[1 + 1.21\left(1 - \sqrt{B/B_{c2}}\right)^3\right]&
\lambda^{-2}[{\rm nm}] .
 \label{eq:sigma_vs_h}
\end{eqnarray}
Here $B$ is the magnetic induction, which for applied fields in
the region $H_{c1}\ll H_{app}\ll H_{c2}$ is $B\simeq \mu_0
H_{app}$, $a$ is the coefficient depending on the symmetry of the
VL ($a=4.83\times10^4$ for triangular VL\cite{Khasanov05,Brandt03}
and, as is shown below, $a=5.07\times10^4$ for rectangular VL),
$H_{c1}$ is the first critical field, and $B_{c2}=\mu_0H_{c2}$ is
the upper critical field. Equation~(\ref{eq:sigma_vs_h}) accounts
for reduction of $\sigma_{sc}$ due to stronger overlapping of
vortices by their cores with increasing field. According to
calculations of Brandt,\cite{Brandt03} Eq.~(\ref{eq:sigma_vs_h})
describes with less than 5\% error the field variation of
$\sigma_{sc}$ for an ideal triangular vortex lattice and it holds
for type-II superconductors with the value of the Ginzburg-Landau
parameter $\kappa=\lambda/\xi\geq5$ ($\xi$ is the coherence
length) in the range of fields $0.25/\kappa^{1.3}\lesssim
B/B_{c2}\leq1$.

Satisfactory fit of Eq.~(\ref{eq:sigma_vs_h}) to the experimental
data would suggest that there is no significant change of the
penetration depth in the range of magnetic fields of the
experiment. We performed an analysis of the magnetic field
dependence of the muon depolarization rate $\sigma_{sc}$ measured
at $T=20$~mK (see Fig.~\ref{fig:lambda_vs_H_muSR}), which is a
good approximation of $\sigma_{sc}(T=0,B)$. The solid line in
Fig.~\ref{fig:lambda_vs_H_muSR} corresponds to the fit of
experimental $\sigma_{sc}(B)$ by means of
Eq.~(\ref{fig:lambda_vs_H_muSR}).  The fit yields
$\lambda(0)=720(25)$~nm and $B_{c2}(0)=1.9(2)$~T. A good agreement
between the experiment and the theory (see
Fig.~\ref{fig:lambda_vs_H_muSR}) allows to conclude that
$\sigma_{sc}(T=0,B)$ can be consistently fitted within the
behavior expected for conventional isotropic superconductor. For
comparison in the inset of Fig.~\ref{fig:lambda_vs_H_muSR} we plot
the dependence of the inverse squared penetration depth normalized
to its value at $B=0$ on the reduced field $b=B/B_{c2}$ for
Mo$_3$Sb$_7$ studied here, the two-gap superconductor MgB$_2$ from
Ref.~\onlinecite{Angst04}, and the hole-doped cuprate
superconductor YBa$_2$Cu$_3$O$_{6.95}$ from
Ref.~\onlinecite{Sonier00}. It is seen that in a case of MgB$_2$
and YBa$_2$Cu$_3$O$_{6.95}$ $\lambda^{-2}$, evaluated from muon
experiments, is field dependent and decreases rather strongly with
increasing field.

\begin{figure}[htb]
\includegraphics[width=0.63\linewidth, angle=-90]{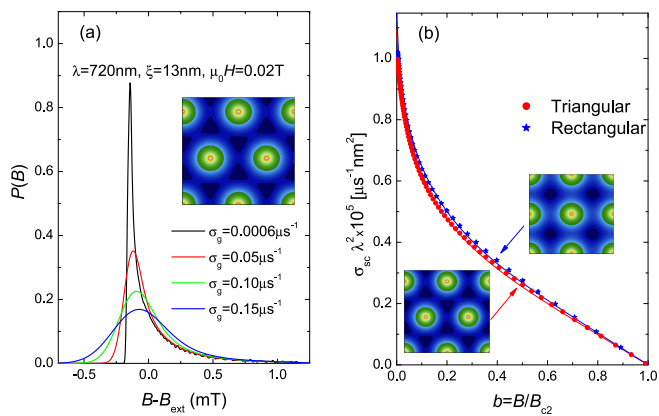}
%
\caption{(Color online) (a) Theoretical $P(B)$ profiles
($\lambda=720$~nm, $\xi=13$~nm, and $\mu_0H=0.02$~T) calculated
for $\sigma_g=0.0006$, 0.05, 0.1, and 0.15~$\mu$s$^{-1}$. (b)
Dependence of the product $\sigma_{sc}\times\lambda^{-2}$ on the
reduced field $b=B/B_{c2}$ for the triangular (circles) and the
rectangular (stars) VL's. Calculations were made by using the
numerical approach of Brandt \cite{Brandt03} for $\kappa=100$. The
solid lines are obtained by means of Eq.~(\ref{eq:sigma_vs_h})
with $a=4.83\cdot10^{4}$ for triangular and
$a=5.07\cdot10^{4}$ for  rectangular VL's, respectively. The insets are the contour plots of field variation within triangular and rectangular VL's. }
 \label{fig:simulations}
\end{figure}

The absolute error in $\lambda(0)$, estimated from the fit of
$\sigma_{sc}(B)$ by means of Eq~(\ref{eq:sigma_vs_h}), is
$\simeq3.5$\%. There are, however, other sources of
uncertainties which could also affect $\lambda(0)$ and,
consequently, increase the error: (i) The use of
Eq.~(\ref{eq:sigma_vs_h}) to extract $\lambda^{-2}$ from
$\sigma_{sc}(B)$ requires that the VL is well ordered and  it is
triangular, as opposed to disordered or rectangular VL.
(ii) The fitting function Eq.~(\ref{eq:Analysis}) assumes that the contribution from the sample is described by a single line of Gaussian shape as opposed to asymmetric magnetic field distribution $P(B)$ generally observed in good quality single crystals.\cite{Sonier00}
(iii) The background signal due to the Ag backing plate was assumed to have a temperature-independent frequency and is non-relaxing at all $T$. However, when the sample goes superconducting it excludes some field, even in the mixed state.  These small fields are very inhomogeneous and could cause as a shift of the background signal to higher fields and an increase of a slow relaxation.
In the following we are going to discuss  in detail the sources
(i) and (ii). We believe that the source (iii) does not play a
substantial role here since even at a $T=20$~mK $B_{int}$ is only
0.16~mT smaller than the applied field and $B_{ext}$ is constant
in the whole temperature range (from 20~mK up to $T_c$). It should be mentioned, however, that experience with more robust superconductors has shown that even though the background field does not usually change perceptibly, the excluded flux can affect the background relaxation rate.

To account for possible random deviations  of the flux core
positions from their ideal ones (VL disorder) and for broadening
of $\mu$SR spectra due to nuclear depolarization, the field distribution of an ideal VL [$P_{id}(B)$] is, generally, convoluted with a Gaussian
distribution in terms of:\cite{Brandt88a,Sonier94,Maisuradze08}
\begin{equation}
P(B)=\frac{1}{\sqrt{2\pi}\sigma_g}\int\exp\left[-\frac{1}{2}\left(
\frac{B-B'}{\sigma_g}\right)^2\right]P_{id}(B')dB'.
 \label{eq:pinning}
\end{equation}
Here $\sigma_g=\sqrt{\sigma_{nm}^2+\sigma_B^2}$ and $\sigma_B$  is
the contribution to the Gaussian broadening of $P_{id}(B)$ due to
VL disorder. The theoretical $P(B)$ profiles
for various $\sigma_g$'s are shown in
Fig.~\ref{fig:simulations}~(a). It is obvious that both, the
nuclear moment contribution and the VL disorder broaden the
$P(B)$ profiles, thus requiring that the total second moment of
$\mu$SR line needs to be the sum of three
components:\cite{Maisuradze08}
$\sigma_{sc}^2+\sigma_{nm}^2+\sigma_{B}^2$. This implies that
neglecting the VL disorder leads to overestimation of
$\sigma_{sc}$ and, as a consequence, to underestimation of
$\lambda$. On the other hand, $\sigma_{sc}$, obtained from the fit
of asymmetric $P(B)$ line by using symmetric Gaussian function [see Eq.~(\ref{eq:Analysis})],
becomes underestimated.\cite{Sonier00}
In a case of Mo$_3$Sb$_7$ the main contribution to $\sigma_{g}$
comes, most probably, from the nuclear moment term
$\sigma_{nm}\simeq0.178$~$\mu$s$^{-1}$, which is comparable with
$\sigma_{sc}$ in the whole field range, rather than from
$\sigma_{B}$, which for good quality single crystals
corresponds, typically, to $10-20$\%, of $\sigma_{sc}$ (see, e.g.,
Ref.~\onlinecite{Sonier00} and references therein). Such big
$\sigma_{nm}$ also leads to the fact that the shape of $P(B)$
profile is close to the Gaussian one as is demonstrated by the
solid blue line in Fig.~\ref{fig:simulations}.

Dependences of $\sigma_{sc}$ on the  reduced field
$b=B/B_{c2}$ for triangular and rectangular VL's, obtained by using numerical calculations of Brandt,
\cite{Brandt03} are shown in
Fig.{\ref{fig:simulations}~(b). Solid lines correspond to Eq.~(\ref{eq:sigma_vs_h}) with $a=4.83\times10^4$ (red line) and
$5.07\times10^4$ (blue line). As is seen, in a case of rectangular VL $\sigma_{sc}(b)$ can still be satisfactory described by means of Eq.~(\ref{eq:sigma_vs_h}) with the only 5\% bigger value of the coefficient $a$. This implies that the uncertainty related to different VL symmetries would result in $\simeq2.5$\% additional error in $\lambda$.
To summarize, by taking into account the  above presented
arguments we believe that the true overall uncertainty in the
absolute value of $\lambda(0)$ is about 10-15~\%, and
$\lambda(0)=720(100)$~nm.

The zero--temperature value of the superconducting coherence
length $\xi(0)$ may be estimated from  $B_{c2}(0)$ as
$\xi(0)=[\Phi_0/2\pi B_{c2}(0)]^{0.5}$, which results in $\xi(0) =
13(1)$~nm ($\Phi_0$ is the magnetic flux quantum). Using the
values of $\lambda(0)$ and $\xi(0)$, the zero-temperature value of
the Ginzburg-Landau parameter is $\kappa(0) = \lambda(0)/\xi(0)=
55(10)$. The value of the first critical field can also be
calculated by means of Eq.~(4) from Ref.~\onlinecite{Brandt03} as
$\mu_0H_{c1}(0) = 1.8(3)$~mT.  Note that the zero-temperature
values of $\lambda(0)$, $\xi(0)$, $\kappa(0)$, and the first and
the second critical fields are in agreement with those
reported in the
literature.\cite{Bukowski02,Dmitriev06,Dmitriev07b,Candolfi08,Tran08}

\begin{figure}[htb]
\includegraphics[width=1.0\linewidth]{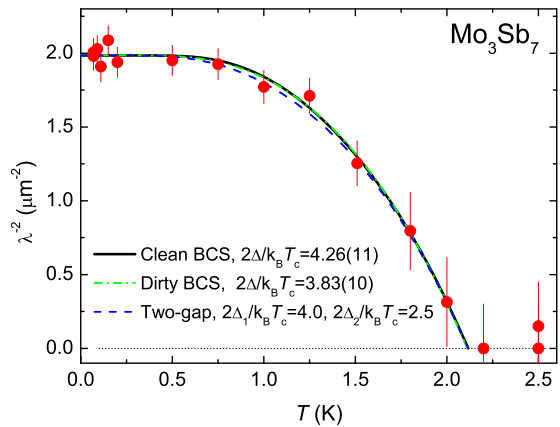}
 \vspace{-1.0cm}
\caption{(Color online) Temperature  dependence of $\lambda^{-2}$
of Mo$_3$Sb$_7$ reconstructed from $\sigma_{sc}(T)$ measured in
$\mu_0H=0.02$~T. The black solid and the dash-dotted blue lines
are the fits by means of clean and dirty weak-coupling BCS models.
The blue dashed line represents
$\lambda^{-2}(T)$ calculated by using two-gap alpha model\cite{Carrington03,Khasanov07_La214} with the parameters
from Ref.~\onlinecite{Tran08}. See text for details.}
 \label{fig:lambda_vs_T}
\end{figure}

As a next step we are going to analyze the temperature dependence
of the magnetic penetration depth. $\lambda^{-2}(T)$, presented in
Fig.~\ref{fig:lambda_vs_T}, was reconstructed from $\sigma_{sc}(T)$
measured in a field of 0.02~T by using Eq.~(\ref{eq:sigma_vs_h}).
$B_{c2}(T)$, needed for such reconstruction, was assumed to follow
the Werthamer-Helfand-Hohenberg prediction\cite{Werthamer66} in
agreement with the results of Ref.~\onlinecite{Candolfi08}. It is worth to mention here that the temperature dependence of $\lambda^{-2}$ is much less affected by the above described uncertainties than the absolute $\lambda(0)$ value. Indeed, both, $\sigma_B$\cite{Sonier94,Maisuradze08} and the relaxation related to the excluded flux increase with decreasing $T$ in much the same way as $\sigma_{sc}$, which causes mainly a correction to the coefficient $a$ in Eq.~(\ref{eq:sigma_vs_h}), while $\sigma_{nm}$ is temperature independent.

Figure \ref{fig:lambda_vs_T} implies that below 0.7~K $\lambda^{-2}$  is {\it temperature
independent} as expected for the superconductor with fully
gaped states.
The experimental $\lambda^{-2}(T)$ dependence was analyzed within
the dirty- and the clean-limit approaches. In the dirty-limit
theory $\lambda^{-2}(T)$ has the form:\cite{Tinkham75}
\begin{equation}
\left. \frac{\lambda^{-2}(T)}{\lambda^{-2}(0)}\right|_{dirty}=
\frac{\Delta(T)}{\Delta(0)}\tanh\left[ \frac{\Delta(T)}{2k_BT}
\right],
 \label{eq:BCS-dirty}
\end{equation}
while in the clean-limit:\cite{Tinkham75}
\begin{equation}
\left.\frac{\lambda^{-2}(T)}{\lambda^{-2}(0)}\right|_{clean}=  1+
2\int_{\Delta(T)}^{\infty}\left(\frac{\partial f}{\partial
E}\right)\frac{E}{\sqrt{E^2-\Delta(T)^2}}\  dE .
 \label{eq:BCS-weak-coupled}
\end{equation}
Here $f=[1+\exp(E/k_BT)]^{-1}$ is  the
Fermi function. The temperature dependence  of the gap was
approximated by
$\Delta(T)=\Delta(0)\tanh\{1.82[1.018(T_c/T-1)]^{0.51}\}$.\cite{Carrington03}
As is seen, both, the dirty- and the clean-limit approaches,
describe the experimental data reasonably well. The fitted curves
are almost undistinguishable from the each other. The results of
the fits are $T_c=$2.11(2)~K, $\Delta(0)=$0.35(1)~meV and
2.12(2)~K, 0.39(1)~meV for the dirty- and the clean-limit BCS
model, respectively. The corresponding gap to $T_c$ ratios were
found to be $[2\Delta(0)/k_BT_c]_{dirty}=3.83(10)$ and
$[2\Delta(0)/k_BT_c]_{clean}=4.27(11)$. There are few reasons why,
we believe, Mo$_3$Sb$_7$ studied here is in the dirty limit:
(i) The maximum gap value obtained in Andreev  reflection
experiments was found to be $\Delta(0)\simeq0.32$~meV which is
close to $\Delta(0)=$0.35(1)~meV observed within our dirty limit
calculations.
(ii) The residual resistivity ratio measured on similar single
crystals  was found to be rather low $\rho_{300{\rm
K}}/\rho_{0{\rm K}}=1.4$.\cite{Bukowski02} Note that the
isostructural compound Ru$_3$Sn$_7$ has $\rho_{300{\rm
K}}/\rho_{0{\rm K}}=144$,\cite{Chakoumakos98} which is bigger by
more than 2 orders of magnitude.
(iii) A reasonable assumption about Fermi velocity
$v_F\simeq10^6$~m/s can  be made within the free electron
approximation by taking $E_F\simeq6.5$~eV.\cite{Dashjav02} This
allows us to readily estimate the BCS-Pippard coherence length
$\xi_0=\hbar v_F/\pi\Delta(0)\simeq600$~nm which is approximately
50 times bigger than $\xi=13(1)$~nm obtained experimentally.
Considering that $\xi^{-1}=\xi_0^{-1}+l^{-1}$ ($l$ is the mean
free path) we obtain that the coherence length $\xi$ in
Mo$_3$Sb$_7$ is limited by $l$ and, correspondingly,
$l\simeq\xi\simeq13$~nm.

Now we are going to comment shortly the results of the specific
heat experiments of Tran {\it et al.}\cite{Tran08} It was shown
that the analysis of the electronic specific heat data within the
framework of the phenomenological alpha model suggests the
presence of two gaps with $2\Delta_1(0)/k_BT_c=4.0$ and
$2\Delta_2(0)/k_BT_c=2.5$ and the relative weight of the bigger
gap of 0.7.\cite{Tran08} The $\lambda^{-2}(T)$ curve simulated by
using these parameters and $\lambda(0)=716$~nm, is shown in
Fig.~\ref{fig:lambda_vs_T} by blue dashed line. Both contributions
were assumed  to be in the dirty limit [see
Eq.~(\ref{eq:BCS-dirty})] and the similar alpha model, but adapted
for calculation of the superfluid density, was
used.\cite{Carrington03,Khasanov07_La214} The agreement between
the simulated curve and the experimental data is rather good. We
should emphasize, however, that in two-gap superconductor the
contribution of the smaller gap to the total superfluid density
decreases very fast with increasing field, thus leading to strong
suppression of
$\lambda^{-2}$.\cite{Serventi04,Angst04,Khasanov07_La214} This is
inconsistent with the data presented in the inset of
Fig.~\ref{fig:lambda_vs_H_muSR} revealing that in Mo$_3$Sb$_7$
$\lambda^{-2}(B)=const$.


To conclude, the
superconductor Mo$_3$Sb$_7$ ($T_c\simeq2.1$~K) was studied by means of muon-spin rotation. The main
results are: (i) The absolute values of the magnetic field
penetration depth $\lambda$, the Ginzburg-Landau parameter
$\kappa$, and the first $H_{c1}$ and the second $H_{c2}$ critical
fields at $T=0$ were found to  be $\lambda(0)=720(100)$~nm,
$\kappa(0)=55(8)$, $\mu_0H_{c1}(0)=1.8(3)$~mT, and
$\mu_0H_{c2}(0)=1.9(2)$~T. (ii) Over the whole temperature range
(from $T_c$ down to 20~mK) the temperature dependence of
$\lambda^{-2}$ is consistent with what is expected for a
single-gap $s-$wave BCS superconductor. (iii) The ratio
$2\Delta(0)/k_BT_c=3.83(10)$ was found, suggesting that
Mo$_3$Sb$_7$ superconductor is in the intermediate-coupling
regime. (iv)
The magnetic penetration depth $\lambda$ is field independent, in
agreement with what is expected for a superconductor with an
isotropic energy gap. (v) The value of the electronic mean-free
path was estimated to be $l\simeq13$~nm. This relatively short
value suggests that strong scattering processes play an important
role in the electronic properties of Mo$_3$Sb$_7$.


This work was performed at the Swiss Muon Source (S$\mu$S),  Paul
Scherrer Institute (PSI, Switzerland). The authors are grateful to
A.~Amato and D.~Herlach for providing the instrumental support
during the $\mu$SR experiments, and A.~Maisuradze for calculating
$\sigma_{sc}(b)$ dependences. The work was supported
by the K.~Alex M\"uller Foundation and in part by the SCOPES grant
No. IB7420-110784.

\end{document}